\documentclass[a4paper]{jpconf}
\usepackage{graphicx}
\begin{document}
\vspace*{-15mm}
\begin{flushright}
{CERN-PH-TH/2007-192}\\
\end{flushright}

\title{Smallness of the cosmological constant and the multiple point principle}

\author{C D Froggatt$^1$, R Nevzorov$^1$ and H B Nielsen$^2$}

\address{$^1$ Department of Physics and Astronomy, Glasgow University, Glasgow G12~8QQ, Scotland}
\address{$^2$ The Niels Bohr Institute, Copenhagen DK~2100, Denmark}

\ead{r.nevzorov@physics.gla.ac.uk}

\begin{abstract}
In this talk we argue that the breakdown of global symmetries in no--scale
supergravity (SUGRA), which ensures the vanishing of the vacuum energy density near the 
physical vacuum, leads to a natural realisation of the multiple point principle (MPP).
In the MPP inspired SUGRA models the cosmological constant is naturally tiny. 
\end{abstract}

\section{No-scale supergravity and the multiple point principle}

In $(N=1)$ supergravity (SUGRA) models the scalar potential is specified in terms of 
the K$\ddot{a}$hler function 
\begin{equation}
G(\phi_{M},\phi_{M}^{*})=K(\phi_{M},\phi_{M}^{*})+\ln|W(\phi_M)|^2\,,
\label{1}
\end{equation}
which is a combination of two functions: K$\ddot{a}$hler potential $K(\phi_{M},\phi_{M}^{*})$
and superpotential $W(\phi_M)$. Here we use standard supergravity mass units: 
$\frac{M_{Pl}}{\sqrt{8\pi}}=1$. In order to break supersymmetry in $(N=1)$ SUGRA models, a hidden sector 
is introduced. It is assumed that the superfields of the hidden sector $(z_i)$ interact with the 
observable ones only by means of gravity. At the minimum of the scalar potential, hidden sector fields
acquire vacuum expectation values breaking local SUSY and generating a non--zero gravitino mass 
$m_{3/2}=<e^{G/2}>$. 

In~~ general~~ the~~ vacuum~~ energy~~ density~~ in~~ SUGRA~~ models~~ is~~ huge~~ and~~ negative~~ 
$\Lambda\sim -m_{3/2}^2 M_{Pl}^2$. The situation changes dramatically in no-scale supergravity 
where the invariance of the Lagrangian under imaginary translations and dilatations results 
in the vanishing of the vacuum energy density. Unfortunately these global symmetries also protect 
supersymmetry which has to be broken in any phenomenologically acceptable theory.

It was argued that the breakdown of dilatation invariance does not necessarily result
in a non--zero vacuum energy density \cite{1}--\cite{2}. This happens if the dilatation invariance 
is broken in the superpotential of the hidden sector only. The hidden sector of the simplest 
SUGRA model of this type involves two singlet superfields, $T$ and $z$, 
that transform differently under dilatations
\begin{equation}
T\to \alpha^2 T\,,\qquad\qquad z\to \alpha z
\label{3}
\end{equation}
The invariance under the global symmetry transformations (\ref{3})
constrains the K$\ddot{a}$hler potential and superpotential of the hidden sector.
In the considered SUGRA model they can be written in the following form \cite{1}:
\begin{equation}
\hat{K}=-3\ln\Biggl[T+\overline{T}-|z|^2 \Biggr]\,,\qquad 
\hat{W}(z)=\kappa\Biggl(z^3+\mu_0 z^2+\sum_{n=4}^{\infty}c_n z^n \Biggr)\,.
\label{5}
\end{equation}
The bilinear mass term for the superfield $z$ and the higher order terms $c_n z^n$ 
in the superpotential $\hat{W}(z)$ spoil the dilatation invariance. However the
SUGRA scalar potential of the hidden sector remains positive definite in the considered model
\begin{equation}
V(T,\, z)=\frac{1}{3(T+\overline{T}-|z|^2)^2}\biggl|\frac{\partial \hat{W}(z)}{\partial z}\biggr|^2\,,
\label{6}
\end{equation}
so that the vacuum energy density vanishes near its global minima. In the simplest case 
when $c_n=0$, the scalar potential (\ref{6}) has two extremum points at $z=0$ and 
$z=-\frac{2\mu_0}{3}$. In the first vacuum where $z=-\frac{2\mu_0}{3}$, local supersymmetry 
is broken and the gravitino gains a non--zero mass. In the second minimum, the vacuum expectation 
value of the superfield $z$ and the gravitino mass vanish. If the high order 
terms $c_n z^n$ are present in Eq.~(\ref{5}), the scalar potential of the hidden sector 
may have many degenerate vacua with broken and unbroken supersymmetry in
which the vacuum energy density vanishes.

Thus the considered breakdown of dilatation invariance leads to a natural realisation of the 
multiple point principle (MPP) assumption. The MPP postulates the existence of the maximal number 
of phases with the same energy density which are allowed by a given theory \cite{3}. 
Successful application of the MPP to $(N=1)$ supergravity requires us to assume the existence of 
a vacuum in which the low--energy limit of the considered theory is described by a pure 
supersymmetric model in flat Minkowski space \cite{4}. According to the MPP this vacuum and the 
physical one in which we live must be degenerate. Such a second vacuum is realised only if the 
SUGRA scalar potential has a minimum where $m_{3/2}=0$ which normally requires an extra fine-tuning 
\cite{4}. In the SUGRA model considered above the MPP conditions are fulfilled automatically without 
any extra fine-tuning.

\section{Cosmological constant in the MPP inspired SUGRA models}

Since the vacuum energy density of supersymmetric states in flat Minkowski space is 
just zero and all vacua in the MPP inspired SUGRA models are degenerate, the 
cosmological constant problem is solved to first approximation by assumption 
in these models. However the value of the cosmological constant may differ from 
zero in the considered models. This occurs if non--perturbative effects in the 
observable sector give rise to the breakdown of supersymmetry in the second vacuum 
(phase). The MPP philosophy then requires that the physical phase in which local supersymmetry 
is broken in the hidden sector has the same energy density as a second phase where 
non--perturbative supersymmetry breakdown takes place in the observable sector.

If supersymmetry breaking takes place in the second vacuum, it is caused by the strong 
interactions. When the gauge couplings at high energies are identical in both vacua and
$M_S$ is the SUSY breaking scale in the physical vacuum the scale $\Lambda_{SQCD}$, 
where the QCD interactions become strong in the second vacuum, is given by
\begin{equation}
\Lambda_{SQCD}=M_{S}\exp\left[{\frac{2\pi}{b_3\alpha_3^{(2)}(M_{S})}}\right]\,,\qquad
\frac{1}{\alpha^{(2)}_3(M_S)}=\frac{1}{\alpha^{(1)}_3(M_Z)}-
\frac{\tilde{b}_3}{4\pi}\ln\frac{M^2_{S}}{M_Z^2}\,. 
\label{8}
\end{equation}
In Eq.(\ref{8}) $\alpha^{(1)}_3$ and $\alpha^{(2)}_3$ are the values of the strong gauge 
couplings in the physical and second minima of the SUGRA scalar potential while 
$\tilde{b}_3=-7$ and $b_3=-3$ are the one--loop beta functions of the SM and MSSM.
At the scale $\Lambda_{SQCD}$ the t--quark Yukawa coupling in the MSSM is of the same 
order of magnitude as the strong gauge coupling. The large Yukawa coupling of the
top quark may result in the formation of a quark condensate that breaks supersymmetry
inducing a non--zero positive value for the cosmological constant
$\Lambda \simeq \Lambda_{SQCD}^4$.

\begin{figure}[h]
\hspace{0cm}{$\log[\Lambda_{SQCD}/M_{Pl}]$}\\
\includegraphics[width=24pc]{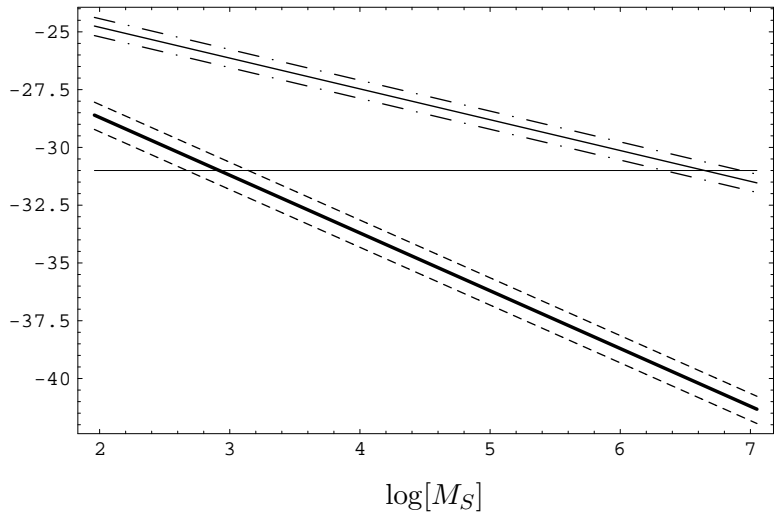}\hspace{2pc}%
\begin{minipage}[b]{12pc}\caption{\label{label}
The value of $\log\left[\Lambda_{SQCD}/M_{Pl}\right]$ versus $\log M_S$.
The thin and thick solid lines correspond to the pure MSSM and the
MSSM with an extra pair of $5+\bar{5}$ multiplets. The dashed and dash--dotted
lines represent the uncertainty in $\alpha_3(M_Z)$, i.e. $\alpha_3(M_Z)=0.112-0.124$.
The horizontal line corresponds to the observed value of $\Lambda^{1/4}$. The SUSY
breaking scale $M_S$ is given in GeV.}
\end{minipage}\\
{\hspace*{5cm}{$\log[M_S]$}}
\end{figure}

In Fig.~1 the dependence of $\Lambda_{SQCD}$ on the SUSY breaking scale $M_S$ is examined. 
When the supersymmetry breaking scale in our vacuum is of the order of 1 TeV, we obtain
$\Lambda_{SQCD}=10^{-26}M_{Pl} \simeq 100$ eV \cite{1}. This results in an enormous suppression 
of the total vacuum energy density ($\Lambda\simeq 10^{-104} M_{Pl}^4$) compared to say 
an electroweak scale contribution in our vacuum $v^4 \simeq 10^{-62} M_{Pl}$.  From 
the rough estimate of the energy density it can be easily seen that the measured value of the
cosmological constant is reproduced when $\Lambda_{SQCD}=10^{-31}M_{Pl} \simeq 10^{-3}$ eV. 
The appropriate values of $\Lambda_{SQCD}$ can therefore only be obtained for 
$M_S=10^3-10^4\,\mbox{TeV}$ \cite{1}. However if the MSSM particle content is supplemented by 
an additional pair of $5+\bar{5}$ multiplets the observed value of the cosmological constant 
can be reproduced even for $M_S\simeq 1\,\mbox{TeV}$ (see Fig.~1). In the physical vacuum these 
extra particles would gain masses around the supersymmetry breaking scale due to the presence 
of the bilinear term $\left[\eta (5\cdot \overline{5})+h.c.\right]$ in $K(\phi_{M},\phi_{M}^{*})$. 
Near the second minimum of the SUGRA scalar potential the new particles would be massless, since 
$m_{3/2}=0$, and would reduce $\Lambda_{SQCD}$ via their contribution to the $\beta$ functions.

\section{Conclusions}
We have shown that the breakdown of global symmetries in no-scale supergravity can lead to a set 
of degenerate vacua with broken and unbroken local supersymmetry (first and second phases) so that 
the MPP conditions are satisfied without any extra fine-tuning. In the MPP inspired SUGRA models 
supersymmetry in the second phase may be broken dynamically in the observable sector inducing
a very small positive energy density which can be assigned, by virtue of the MPP, to all other phases. 
In such a way we have suggested an explanation of why the observed value of the cosmological constant 
is positive and takes on the tiny value it has.

\ack
RN acknowledge support from the SHEFC grant HR03020 SUPA 36878.

\end{document}